\documentclass[12pt]{article}
\usepackage[dvips] {graphics}
\usepackage {epsfig}
\begin{document}
\baselineskip=0.7cm

\overfullrule=0pt
\centerline{\bf DETECTION OF INCOMPATIBLE PROPERTIES}
\par\centerline{\bf WITHOUT ERASURE}
\vskip1.2cm
\centerline{Giuseppe Nistic\`o}
\par
\centerline{{\it Dipartimento di Matematica -
Universit\`a della Calabria}, and}
\par
\centerline{\it Istituto Nazionale di Fisica Nucleare - Gruppo coll. di Cosenza}
\par
\centerline{87036 Arcavacata, Rende (CS), Italy}
\par
\centerline{gnistico@unical.it}
\vskip2pc\noindent {\bf Abstract.}\quad
In this work we show that in double-slit experiment properties incompatible with Which Slit property can
be detected without erasing the knowledge of which slit each particle passes through
and without affecting the point of impact on the final screen.
A systematic procedure to find these particular properties is provided.
A thought experiment which realizes this detection is proposed.
\vfill\eject\noindent
\centerline{\bf INTRODUCTION}
\vskip1pc\noindent
Since the birth of Quantum Mechanics, the double-slit experiment has been used
in the literature for its effectiveness in
illustrating the duality between corpuscolar and wave-like
behaviours of physical entities [1-3].
In relatively recent works,
Englert, Scully and Walther (ESW) were able to push
the understanding of the origin of complementarity and of its physical
implications to a deeper level [4,5], by using just an ideal double-slit experiment for atoms.
A particular set-up of the experiment designed by ESW makes it possible
to detect which slit each particle passes through,
{\sl without essentially perturbing the momentum of the particle}; this notwithstanding, no interference
appears on the final screen.
This result led the quoted authors to conclude that the mutual exclusion
between the particle-like property ``{\sl passing through one of the slits}'' ({\sl Which Slit property})
and the wave-like property
``{\sl emergence of interference pattern on the final screen}'' cannot be explained
as usual by arguing that the act of observation,
while the particle passes through a slit, scatters its momentum by an amount sufficient to destroy the
interference pattern (see [6-11] for the debate about this point, which is not the subject of the present work).
A different set-up of ESW experiment makes
possible the non-disturbing detection of another property of the system,
{\sl incompatible} with Which Slit property; in so doing interference is restored but the knowledge of which slit the particle passes through is definitively lost (erased): this phenomenon has been called {\sl erasure}.
In the present work we show, by means of a systematic theoretical investigation, how it is possible to detect
properties incompatible with Which Slit property without erasing Which Slit knowledge and without affecting the point of 
impact on the final screen.
The remaining part of
this introduction is devoted to explain the ideas our work is based on and to outline
the main results.\par
In ESW experiment the non-disturbing detection is made possible by the fact that
besides the position of the centre-of-mass of the atom, the system possesses further degrees of freedom.
As a consequence, the Hilbert space describing the entire system can be decomposed as
${\cal H}={\cal H}_1\otimes{\cal H}_2$, where ${\cal H}_1$ is the Hilbert space used to represent
the observable {\sl position} (of the centre-of-mass of the atom), its conjugate momentum,
and all other observables arising from them.
An example of this first kind of observables is just {\sl Which Slit property}, we define as the observable which
takes values 1 or 0 according to whether the atom is observed to pass through slit 1 or slit 2. It is represented by a
projection operator $E$ which acts on ${\cal H}_1$.
Hilbert space ${\cal H}_2$ is used to represent the observables arising from the further degrees of freedom,
related to spin or similar.
The non-disturbing detection of Which Slit property is obtained by measuring an observable of this second type,
represented by a particular projection operator $T$ which acts on ${\cal H}_2$,  whose outcomes are correlated with the outcomes of $E$:
outcome 1 (resp., 0) of $T$ reveals the passage through slit 1 (resp., 2).
Since in ESW experiment the Hamiltonian operator $H$ is essentially independent of the further degrees of freedom,
the Which Slit detection by means of $T$ does not affect
the centre-of-mass motion of the particle.
\par
To obtain {\sl erasure},
ESW introduce another property, represented by a projection operator $E_+$ acting on ${\cal H}_1$, incompatible with Which Slit property, i.e. such that
$[E,E_+]\neq {\bf 0}$.
Also such a new property can be detected without affecting the centre-of-mass motion of the atom, by means of another
non-disturbing detector $T_+$ acting on ${\cal H}_2$, which plays with respect to $E_+$ the same role played by $T$ with respect to $E$.
However, the two non-disturbing detections cannot be performed together because
$[T,T_+]\neq{\bf 0}$.
Furthermore, if $T_+$ is applied, and we select the atoms which turn out to possess property $E_+$,
i.e. such that the outcome of $T_+$ is 1, then their distribution exhibits interference.
The presence of interference implies the impossibility, according to standard quantum theory, of assigning Which Slit property to each atom (see section V).
Thus, in the particular example studied by ESW, the non-disturbing detection of $E_+$ erases WS property.
\par
This result, however, does not erase the possibility of devising experimental situations for detecting a
property incompatible with Which Slit property without erasing
Which Slit knowledge and without disturbing the centre-of-mass motion.
We envisage a situation in which the second property, incompatible with Which Slit property, can be detected
by means of a projection operator $Y$ acting on ${\cal H}_2$, but which {\sl can be measured together with} $T$.
More precisely, our aim is to find a concrete Hilbert Space ${\cal H}={\cal H}_1\otimes{\cal
H}_2$ for describing a double-slit experiment, where Which Slit
property is represented by a projection operator $E$ which acts on
${\cal H}_1$, such that a concrete state vector $\Psi$ and a
concrete projection operator $G$ representing a property
incompatible with $E$ (i.e. $[E,G]\neq{\bf 0}$) can be singled out
for which
\begin{description}
\item[{\rm a)}]
Which Slit property $E$ can be detected by a non-disturbing detector $T$ acting on ${\cal H}_2$.
\item[{\rm b)}]
Property $G$ can be detected by another non-disturbing detector $Y$ acting on ${\cal H}_2$
\item[{\rm c)}]
The two non-disturbing detections can be carried out together, i.e. $[Y,T]={\bf 0}$,
so that detecting $G$ does not erase Which Slit property.
\end{description}
We have found that the success of such a research strongly depends
on the dimension of space ${\cal H}_1$. The non-disturbing
detection of a property incompatible with Which Slit property is
possible without erasing this last only if {\it dim}$({\cal
H}_1)\geq 4$.
\par
In sect. I we establish the theoretical apparatus for dealing with the problem at issue.
Within the general mathematical formalism we introduce the notion of {\sl non-disturbing} detector
of a given property as a generalization of the concept of non-disturing Which Slit
detector devised by ESW.
\par
The problem of the possibility of detecting a property
incompatible with Which Slit property without erasing the latter and without affecting the point of the final impact
is formulated in formal way as problem ($\cal P$) in section II. To
make easier the search of solutions of the problem, a particular
matrix representation is adopted in section II.A. In this
representation it is possible to concretely find solutions whenever
four general constraints, we named (gc.1)-(gc.4) are satisfied.
\par
We begin a systematic research of solutions, by considering the
minimum dimension of ${\cal H}_1$, i.e. {\it dim}$({\cal H}_1)=2$
(sect. II.C).
No solution exists in this case:
whenever non-disturbing detection of 
a property incompatible with Which Slit property is available,
Which Slit knowledge is erased, as it happens in the particular experimental situation put forward by ESW.
\par
If ${\cal H}_1$ has dimension 4 (we consider only even dimensions
to have symmetrical slits), there are properties, incompatible
with Which Slit property, which are detectable without erasing
Which Slit property; concrete solutions are found in section III.A and III.B.
But the two detections turn out to be always correlated, e.g.
a particle is revealed to possess the incompatible
property if and only if it is detected to pass through slit 1.
\par
This last result leads to the question whether non-correlated solutions do exist or not.
This question is affirmatively answered in section IV, where the case {\it dim}$({\cal H}_1)\geq 6$ is investigated.
We find that both correlated and non-correlated solutions exist, and
families of non-correlated solutions are concretely singled out.
\par
Section V is devoted to see what happens to interference when
non-disturbing detections are available. Within our formalism, we
quite simply prove that {\sl no interference can resort for the
particles selected by means of a non-disturbing detector which
does not erase Which Slit property (no-interference theorem)}.
This result agrees with the Wave-Particle duality relation
$V^2+K^2\leq 1$ between the visibility $V$ of interference and and
the knowledge $K$ of which slit the particle passes through
[12,13].
\par
Though our results are established on a theoretical ground, in the conclusive section VI we suggest
an ideal experiment for realizing Which Slit detection together with an incompatible property, without affecting
the distribution of particles on the final screen. The experimental set-up proposed by us, which
exploits the physical properties on micromaser cavities as ESW experiment does, corresponds to a non-correlated
theoretical solution singled out in section IV.
\vskip1.2cm\noindent
{\bf I. SIMULTANEOUS DETECTION OF INCOMPATIBLE PROPERTIES}
\vskip1pc\noindent
We introduce the mathematical quantum formalism to describe
a typical two-slit experiment,
which allows non-disturbing detection. The physical system consists of a localizable particle
whose position (of the centre-of-mass) observable is represented, at time $t$ in Heisenberg picture,
by an operator $Q^{(t)}$ of a suitable Hilbert space ${\cal H}_1$. Let our particle
be endowed with further degrees of freedom, related to
spin or similar, which are described in a different
Hilbert space ${\cal H}_2$, in such a way that the complete Hilbert space is ${\cal H}={\cal H}_1\otimes {\cal H}_2$.
Let us suppose that the dependence of the
Hamiltonian operator $H$ on the degrees of freedom described in ${\cal H}_2$ can be neglected, so that we may assume the ideal case
that $H=H_1\otimes {\bf 1}_2$, where $H_1$ is a self-adjoint operator of ${\cal H}_1$.
\par
To simplify the formalism, without losing generality, we can assume that
the projection operator representing the property
``the particle passes through slit 1 (resp., 2)'' has the form
$E=L\otimes{\bf 1}_2$ (resp., $E'\equiv({\bf 1}-E)=({\bf 1}_1-L)\otimes{\bf 1}_2$),
where $L$ is the localization projection operator of ${\cal H}_1$ which localizes the particle in slit 1.
We shall identify $E$ as {\sl Which Slit} (WS) property.
Given any interval $\Delta$ on the final screen,
by $F(\Delta)$ we denote the projection operator which represents the event
``the particle hits the final screen in a point within $\Delta$''. Also $F(\Delta)$, like $E$, is a localization operator, so that it must have the form $F(\Delta)=J\otimes{\bf 1}_2$, where $J$ is a localization projection operator of ${\cal H}_1$ as $L$, but corresponding to the time, say $t_2$, of the final impact whereas $L$ is relative to the time, say $t_1$, at which the particle crosses the screen that supports the slits. Since rule
$[H_1,Q^{(t)}]={\bf 0}_1$ in general does not hold, we cannot assume $[L,J]={\bf 0}_1$.
Therefore,
it is not generally possible to perform a measurement
revealing which slit is the way of a particle hitting in $\Delta$.
\vskip1pc\noindent
\centerline{\bf A. Which Slit detector}
\vskip.5pc\noindent
However, in particular cases the detection of which slit the particle passes through is possible for each particle hitting the final screen,
without essentially affecting its centre-of-mass motion [4,5].
This detection may happen if in correspondence with the state vector $\Psi$ describing the system
a projection operator of the kind $T={\bf 1}_1\otimes S$ exists such that
equation $T\Psi=E\Psi$ holds.
Projection $T$ represents a property linked to the degrees of freedom described in ${\cal H}_2$; but,
since $[T,E]={\bf 0}$, equation $T\Psi=E\Psi$ is equivalent to say that both conditional probabilities
$$
p(T\mid E)={\langle \Psi\mid TE\Psi\rangle\over\langle\Psi\mid E\Psi\rangle}\,,\quad p(E\mid T)={\langle \Psi\mid TE\Psi\rangle\over\langle\Psi\mid T\Psi\rangle}
$$
are equal to 1, so that the occurrence of outcome 1 (resp., 0) for $T$ detects the passage of the particle through slit 1 (resp.2).
Furthermore,
the outcome of $T$ can be ascertained together with the point of the impact and without affecting
the distribution
of the particles on the final screen, because $[T,F(\Delta)]={\bf 0}$.
Therefore,
such a measurement of $T$ provides WS knowledge without disturbing
the centre-of-mass motion of the particle, so that projection $T$ can be called
WS (non-disturbing) detector.
\begin{figure}[htb!]
\centerline{\epsfig{figure=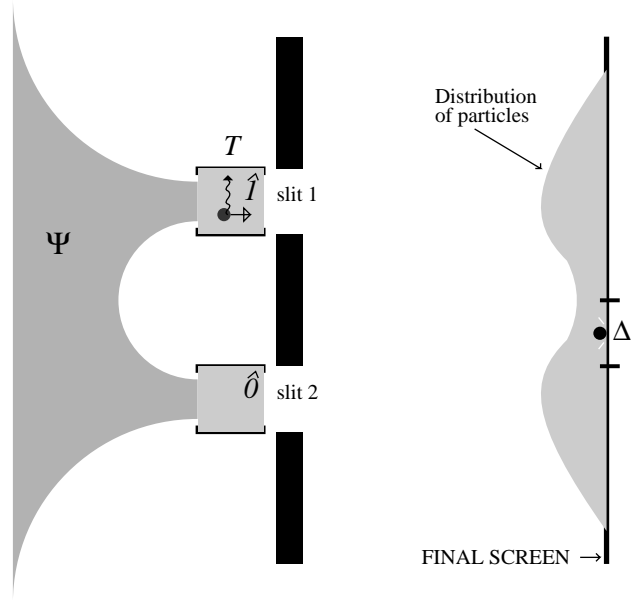,height=8cm}}
\caption[]{Which Slit detector} \label{fig1}
\end{figure}

\vskip1pc\noindent {\bf Example 1}. \quad The thought experiment
of ESW is a particularly effective example of the general scheme
just outlined. The physical system
consists of an atom
in a long lived excited state, e.g. {\it rubidium} in state $63p_{3/2}$.
Hilbert space ${\cal H}_1$ is taken as
the Hilbert space describing the centre-of-mass motion of the
atom. The atom is prepared to travel towards the two slits,
but non elsewhere. The further degrees of freedom, to be described in ${\cal H}_2$, concern with
a pair of cavities $\hat 1$ and $\hat 0$,
placed as shown in fig. 1.
The cavities are resonators for the
electromagnetic field, tuned at a microwave frequency in such a
way that whenever the excited atom enters cavity $\hat 1$ or $\hat
0$, it decays emitting a photon. Whether the photon is emitted in
cavity $\hat 1$ or $\hat 0$ is ascertained by means of suitable
devices the cavities are equipped with. The event ``a photon is
revealed in cavity $\hat 1$ (resp., $\hat 0$)'' is represented by
a projection operator $S=\vert 1\rangle\langle 1\vert$ (resp.,
$S'=\vert 0\rangle\langle 0\vert $) of a Hilbert space ${\cal
H}_2$, describing the pair of cavities and different from
${\cal H}_1$. Here $\vert 1\rangle$, $\vert 0\rangle$  are state
vectors of ${\cal H}_2$ such that $S\vert 1\rangle =\vert
1\rangle$ and $S'\vert 0\rangle=\vert 0\rangle$.
\par
In this experimental situation the complete state vector of the
particle is $\Psi={1\over\sqrt 2}[\psi_1\otimes\vert
1\rangle+\psi_2\otimes\vert 0\rangle]$, where
$\psi_1,\psi_2\in{\cal H}_1$ are state vectors respectively
localized in slit 1 and 2 when the particle is in the region of
the two-slit, i.e. $L\psi_1=\psi_1$, $L\psi_2=0$ but such that
${\it Re}\langle\psi_1\mid J\psi_2\rangle\neq 0$ (owing to
$[L,J]\neq{\bf 0}$). A non-disturbing WS detector is represented
by the projection operator $T={\bf 1}_1\otimes\vert
1\rangle\langle 1\vert$; indeed $({\bf 1}_1\otimes\vert
1\rangle\langle 1\vert)\Psi=(L\otimes{\bf 1}_2)\Psi$, i.e. $T\Psi=E\Psi$, trivially
holds. It must be noticed that in this case the Hamiltonian $H$ of
the entire system depend on $S$. Indeed, the emission of a photon
causes a change in the momentum of the atom. However, it has been shown [14]
that such a dependence can be neglected, as required,
because the photon energy
is very low with respect to that of the atom.
In particular, the absence of interference cannot be caused by a stochastic transfer of momentum between
the atoms and the WS detector.
\vskip1pc
\centerline{\bf B. Non-disturbing detectors} \vskip.5pc\noindent
The concept of non-disturbing WS detector can be extended to
include non-disturbing detector of more general properties of the
kind $G=K\otimes{\bf 1}_2$, according to the following definition.
\vskip1pc\noindent {\bf Definition 1}. {\sl Let $K$ be a
projection operator of ${\cal H}_1$. A projection operator $Y$ of
$\cal H$ is called a non-disturbing detector of property
$G=K\otimes{\bf 1}_2$ with respect to the state vector $\Psi$ if
\par\noindent
i)\quad $[Y,F(\Delta)]={\bf 0}$\par\noindent
ii)\quad $[Y,G]={\bf 0}\;$ and $\;Y\Psi=G\Psi$.}
\par\noindent
Condition (i) states that a measurement of $Y$ can be performed without affecting the distribution of the particles on the final screen. Condition (ii) ensures that $Y$ plays, with respect to $G$, the same role played by WS detector $T$ with respect to WS property $E$, so that the occurrence of outcome 1 (resp., 0) 
for Y reveals that the particle has the property $G$ (resp., $G'\equiv {\bf 1}-G$) [15].
\vskip1pc
It is possible that with respect to
the same state vector $\Psi$, besides a non-disturbing
WS detector $T$, there is a non-disturbing detector
$Y$ of a property $G=K\otimes{\bf 1}_2$ which is incompatible with WS property $E=L\otimes{\bf 1}_2$, i.e.
such that $[E,G]\neq{\bf 0}$.
\vskip1pc\noindent
{\bf Example 2}.
Making reference to example 1,
with respect to $\Psi={1\over\sqrt 2}[\psi_1\otimes\vert 1\rangle+\psi_2\otimes\vert 0\rangle]$,
besides the non-disturbing WS detector
$T={\bf 1}_1\otimes\vert 1\rangle\langle 1\vert$ (of property
$E=\vert\psi_1\rangle\langle\psi_1\vert\otimes{\bf 1}_2$),
there is a non-disturbing detector $T_+={\bf 1}_1\otimes\vert+\rangle\langle+\vert$ of property
$\vert\psi_+\rangle\langle\psi_+\vert\otimes{\bf 1}_2$,
where $\psi_+=1/\sqrt 2(\psi_1+
\psi_2)$ and $\vert +\rangle=(1/\sqrt 2)(\vert 1\rangle+\vert 0\rangle)$.
In this case,
the two properties $E$ and $E_+$ do not commute: $[E,E_+]\neq{\bf 0}$.
The non-disturbing detectors $T$ and $T_+$ cannot be used on the same system because $[T,T_+]\neq{\bf 0}$.
Therefore, only an alternative, mutually exclusive choice can be done between the measurements of
$T$ and $T_+$.
As shown by ESW, if $T_+$ is chosen, we detect $E_+$, but the WS knowledge which would have provided a measurement
of $T$ is erased [4,5].
\vskip1pc
\centerline{\bf II. GENERAL CONSTRAINTS}
\vskip1pc\noindent
In the present work we seek, on a theoretical ground, for the possibility of detecting a property $G=K\otimes{\bf 1}_2$ incompatible with WS property $E$, but {\sl without erasing WS
knowledge provided by a non-disturbing WS detector $T$},
and without affecting the distribution of the particles on the final screen.
This possibility can be realized if
with respect to the same state vector $\Psi$ there exist both a WS non-disturbing detector $T$ of $E$ and a
non-disturbing detector $Y$ of $G$; moreover, it must be required that $[Y,T]={\bf 0}$, so that
$Y$ and $T$ can be measured together, yielding simultaneous information about $E$ and $G$.
In this section these requirements are formulated as general constraints for $K$ and $\Psi$ within a representation particularly suitable for the searching of solutions.
\par
If $G$ is detected by means of a detector of the kind $Y={\bf 1}_1\otimes R$,
then the distribution of the particles on the final screen is not affected
because $[Y,F(\Delta)]={\bf 0}$.
Therefore, our task will be successful if the following problem has solution.
(Henceforth, by abuse of language, we denote the identity operator and the null operator by
${\bf 1}$ and ${\bf 0}$, without specifying the Hilbert space they act on, because such a space turns out to be
self-evident, so that no confusion is likely).

\vskip1pc\noindent
(${\cal P}$)\quad
{\sl
Given WS property $E=L\otimes{\bf 1}$, we have to find
\par\noindent
-- a projection operator $K$ of ${\cal H}_1$, to form $G=K\otimes{\bf 1}$,
\par\noindent
--
a projection operator $S$ of ${\cal H}_2$, to form $T={\bf 1}\otimes S$,
\par\noindent
--
a projection operator $R$ of ${\cal H}_2$, to form $Y={\bf 1}\otimes R$,
\par\noindent
--
a state vector $\Psi\in{\cal H}_1\otimes{\cal H}_2$,
\par\noindent
such that the following conditions hold
\par\noindent
(C.1)\quad
$[E,G]\neq{\bf 0}$, i.e $[L,K]\neq{\bf 0}$,
\par\noindent
(C.2)\quad
$[T,Y]={\bf 0}$, i.e $[S,R]={\bf 0}$,
\par\noindent(C.3)\quad
$T\Psi=E\Psi$,
\par\noindent
(C.4)\quad $Y\Psi=G\Psi$,
\par\noindent
(C.5)\quad
$0\neq E\Psi\neq\Psi$, $0\neq G\Psi\neq \Psi$.}
\vskip1pc
\noindent
(C.5) is added to exclude solutions of (C.1)-(C.4) corresponding to the uninteresting case
that $\Psi$ is eigenvector of $E$ or $G$.
Conditions (C.1)-(C.4) can be more effectively expressed as (gc.1)-(gc.4) below, if the
matrix representation we introduce in sub-section A is adopted.
\vskip1pc
\centerline{\bf A. Matrix representation}
\vskip.5pc\noindent
{\sl Representation for ${\cal H}_1$}.\quad
Let $\{e_1,e_2,e_3,...;r_1,r_2,r_3,...\}$ be an orthonormal basis of ${\cal H}_1$,
made up of eigenvectors of $L$, such that $Le_k =e_k$ and
$Lr_j=0$ for all $k$ and $j$.
Every vector $\psi\in {\cal H}_1$ shall be represented as the column vector
$\psi=\left[\matrix{\varphi_1\cr
{\varphi_0}\cr}\right]$, where
${\varphi_1}=\left[\matrix{\langle e_1\mid \psi\rangle\cr
\langle e_2\mid \psi\rangle\cr
\cdot\cr}\right]$ and ${\varphi_0}=\left[\matrix{
\langle r_1\mid \psi\rangle
\cr
\langle r_2\mid\psi\rangle\cr
\cdot}\right]$.
In such a representation,
every linear operator $W:{\cal H}_1\to{\cal H}_1$ is represented as a matrix
$\left[\matrix{P&U\cr V&Q\cr}\right]$, where $P$, $U$, $V$ and $Q$ are matrices such that
$$
W\psi=\left[\matrix{P&U\cr V&Q\cr}\right]\left[\matrix{\varphi_1\cr {\varphi_0}\cr}\right]
=\left[\matrix{P\varphi_1+U\varphi_0\cr V\varphi_1+Q\varphi_0}\right]\,.
$$
Therefore, projections operators $L$ and $K$ in (C.1)-(C.5) must have the following representations.
$$
L=\left[\matrix{{\bf 1}& {\bf 0}\cr {\bf 0}&{\bf 0}\cr}\right]\;,\quad
K=\left[\matrix{P&U\cr V&Q\cr}\right]\,,\hbox{ where }\;U\neq {\bf 0}\,,K=K^\ast=K^2.\eqno(gc.1)
$$
Constraint $U\neq{\bf 0}$ above is equivalent to
$[L,K]\neq {\bf 0}$ required by (C.1). So, we have established the first general constraint.
\vskip1pc\noindent
{\sl Representation for ${\cal H}_2$}.\quad
Condition (C.2) implies that four projection operators
$A$, $B$, $C$, $D$ of ${\cal H}_2$ exist such that
$A+B+C+D={\bf 1}  $,
$S=A+B$ and $R=A+C$ [16]. Then we choose to represent
every vector ${\bf x}\in{\cal H}_2$ as a column vector
$$
{\bf x}=[a,b,c,d]^T\,\;\hbox{ where }\;{ a}=A{\bf x}\,,\;
{ b}=B{\bf x}\,,\;{ c}=C{\bf x}\,,\;{ d}=D{\bf x}\,.
\eqno(1)
$$
As a consequence, every linear operator $X:{\cal H}_2\to{\cal H}_2$
is represented as a matrix
$X=\left[\matrix{X_{AA}&X_{AB}&X_{AC}&X_{AD}\cr X_{BA}&X_{BB}&X_{BC}&X_{BD}\cr
X_{CA}&X_{CB}&X_{CC}&X_{CD}\cr X_{DA}&X_{DB}&X_{DC}&X_{DD}\cr}\right]$
whose elements are the linear mappings (e.g. $X_{BC}:C{\cal H}_2\to B{\cal H}_2$) such that
$$
X{\bf x}=\left[\matrix{X_{AA}{ a}+X_{AB}{ b}+X_{AC}{ c}+X_{AD}{ d}\cr X_{BA}{ a}+X_{BB}{ b}+X_{BC}{ c}+X_{BD}{ d}\cr
X_{CA}{ a}+X_{CB}{ b}+X_{CC}{ c}+X_{CD}{ d}\cr X_{DA}{ a}+X_{DB}{ b}+X_{DC}{ c}+X_{DD}{ d}\cr}\right].
\eqno(2)
$$
Therefore, projection operators $S$ and $R$ in (C.1)-(C.5) must satisfy the following constraints.
$$
S=A+B=\left[\matrix{{\bf 1}&{\bf 0}&{\bf 0}&{\bf 0}\cr {\bf 0}&{\bf 1}&{\bf 0}&{\bf 0}\cr
{\bf 0}&{\bf 0}&{\bf 0}&{\bf 0}\cr {\bf 0}&{\bf 0}&{\bf 0}&{\bf 0}\cr}\right]\;,\quad R=A+C=
\left[\matrix{{\bf 1}&{\bf 0}&{\bf 0}&{\bf 0}\cr {\bf 0}&{\bf 0}&{\bf 0}&{\bf 0}\cr
{\bf 0}&{\bf 0}&{\bf 1}&{\bf 0}\cr {\bf 0}&{\bf 0}&{\bf 0}&{\bf 0}\cr}\right]
\;.
\eqno(gc.2)
$$
\vskip1pc\noindent
{\sl Representation for ${\cal H}={\cal H}_1\otimes {\cal H}_2$}.\quad
Once fixed the basis $\{e_1,e_2,...;\; r_1,r_2,...\}$,
every vector $\Psi\in{\cal H}_1\otimes{\cal H}_2$ can be uniquely decomposed as
$\Psi=\sum_j{\bf e}_j\otimes{\bf x}_j+\sum_k{\bf r}_k\otimes{\bf y}_k$,
where ${\bf x}_j,{\bf y}_k\in{\cal H}_2$.
Then $\Psi$ shall be represented as a column vector
$$
\Psi=\left[\matrix{{\bf x}_1\cr{\bf x}_2\cr\cdot\cr{\bf y}_1\cr{\bf y}_2\cr\cdot\cr\cdot}\right]=
[\underbrace{{a}_1,{b}_1,{c}_1,{d}_1}_{{\bf x}_1^T},\underbrace{{a}_2,{b}_2,{c}_2,{d}_2}_{{\bf x}_2^T},
\cdot,\cdot;\;\underbrace{\alpha_1,\beta_1,\gamma_1,\delta_1}_{{\bf y}_1^T},
\underbrace{\alpha_2,\beta_2,\gamma_2,\delta_2}_{{\bf y}_2^T},\cdot,\cdot]^T\,,
$$
where, according to (1), $a_j=A{\bf x}_j$, $b_j=B{\bf x}_j$, $c_j=C{\bf x}_j$, $d_j=D{\bf x}_j$ and $\alpha_k=A{\bf y}_k$, $\beta_k=B{\bf y}_k$, $\gamma_k=C{\bf y}_k$,
$\delta_k=D{\bf y}_k$.
If $W=\left[\matrix{P&U\cr V&Q\cr}\right]$ is a linear operator of ${\cal H}_1$ and $X$ is a linear operator of ${\cal H}_2$,
the linear operator $W\otimes X$ of ${\cal H}={\cal H}_1\otimes{\cal H}_2$ is representend as the matrix
$$
W\otimes X=
\left[\matrix{p_{11}X&p_{12}X&\cdot&u_{11}X&u_{12}X&\cdot\cr
              p_{21}X&p_{22}X&\cdot&u_{21}X&u_{22}X&\cdot\cr
              \cdot  & \cdot &\cdot&\cdot  &\cdot  &\cdot\cr
              v_{11}X&v_{12}X&\cdot&q_{11}X&q_{12}X&\cdot\cr
              v_{21}X&v_{22}X&\cdot&q_{21}X&q_{22}X&\cdot\cr
              \cdot  & \cdot &\cdot&\cdot  &\cdot  &\cdot\cr
              }\right]
\eqno(3)
$$
such that
$$
(W\otimes X)\Psi
=\left[\matrix{p_{11}X&p_{12}X&\cdot&u_{11}X&u_{12}X&\cdot\cr
              p_{21}X&p_{22}X&\cdot&u_{21}X&u_{22}X&\cdot\cr
              \cdot  & \cdot &\cdot&\cdot  &\cdot  &\cdot\cr
              v_{11}X&v_{12}X&\cdot&q_{11}X&q_{12}X&\cdot\cr
              v_{21}X&v_{22}X&\cdot&q_{21}X&q_{22}X&\cdot\cr
              \cdot  & \cdot &\cdot&\cdot  &\cdot  &\cdot\cr
              }\right]
\left[\matrix{{\bf x}_1\cr{\bf x}_2\cr\cdot\cr{\bf y}_1\cr{\bf y}_2\cr\cdot}\right]=
\left[\matrix{\sum_j p_{1j}X{\bf x}_j+\sum_k u_{1k}X{\bf y}_k\cr
              \sum_j p_{2j}X{\bf x}_j+\sum_k u_{2k}X{\bf y}_k\cr
              \cdot\cr
              \sum_j v_{1j}X{\bf x}_j+\sum_k q_{1k}X{\bf y}_k\cr
              \sum_j v_{2j}X{\bf x}_j+\sum_k q_{2k}X{\bf y}_k\cr
              \cdot\cr
              }\right]\eqno(4)
$$
\vskip1pc
\centerline{\bf B. Constraints for $\Psi$ and $K$}
\vskip.5pc\noindent
Now we establish which form $\Psi$ must have to satisfy (C.3).
According to (gc.1) and (3), WS property $E=L\otimes{\bf 1}$
and its non-disturbing detector $T={\bf 1}\otimes S$
are represented as
$$
E=
\left[\matrix{{\bf 1}&  {\bf 0}     &\cdot&{\bf 0}  &{\bf 0}  &\cdot\cr
              {\bf 0}&{\bf 1}    &\cdot& {\bf 0}   &{\bf 0}    &\cdot\cr
              \cdot & \cdot &\cdot&\cdot  &\cdot  &\cdot\cr
               {\bf 0}  & {\bf 0} &\cdot&  {\bf 0}  &   {\bf 0}     &\cdot\cr
                 {\bf 0}    &  {\bf 0}      &\cdot&  {\bf 0}      &  {\bf 0}      &\cdot\cr
              \cdot  & \cdot &\cdot&\cdot  &\cdot  &\cdot\cr
              }\right]
,\quad T=
\left[\matrix{S &{\bf 0} &\cdot&{\bf 0} &{\bf 0} &\cdot\cr
              {\bf 0} & S&\cdot&{\bf 0} &{\bf 0} &\cdot\cr
              \cdot & \cdot &\cdot&\cdot  &\cdot  &\cdot\cr
              {\bf 0} &{\bf 0} &\cdot& S &{\bf 0} &\cdot\cr
              {\bf 0} & {\bf 0}&\cdot&{\bf 0} &S &\cdot\cr
              \cdot  & \cdot &\cdot&\cdot  &\cdot  &\cdot\cr
              }\right].\eqno(5)
$$
If $\Psi=[{\bf x}_1,{\bf x}_2,\cdots,{\bf x}_j;\;\cdots,{\bf y}_1,{\bf y}_2,\cdots,{\bf y}_k,\cdots]^T$, then
condition (C.3) $T\Psi=E\Psi$ , with (4) and (gc.2), implies $c_j=d_j=0$, $\alpha_k=\beta_k=0$ i.e.
$$
{\bf x}_j=\left[\matrix{a_j\cr b_j\cr 0\cr 0}\right]\,,\quad {\bf y}_k=\left[\matrix{0\cr 0\cr \gamma_k\cr\delta_k}\right],
\forall j,k\,.\eqno(gc.3)
$$
Further constraints are imposed by (C.4). If $K=\left[\matrix{P&U\cr V&Q\cr}\right]$, then according to
(3) and (gc.3) we have that $G=K\otimes{\bf 1}$ and $Y={\bf 1}\otimes R$ are represented as
$$
G=\left[\matrix{p_{11}{\bf 1}&p_{12}{\bf 1}&\cdot&u_{11}{\bf 1}&u_{12}{\bf 1}&\cdot\cr
              p_{21}{\bf 1}&p_{22}{\bf 1}&\cdot&u_{21}{\bf 1}&u_{22}{\bf 1}&\cdot\cr
              \cdot  & \cdot &\cdot&\cdot  &\cdot  &\cdot\cr
              v_{11}{\bf 1}&v_{12}{\bf 1}&\cdot&q_{11}{\bf 1}&q_{12}{\bf 1}&\cdot\cr
              v_{21}{\bf 1}&v_{22}{\bf 1}&\cdot&q_{21}{\bf 1}&q_{22}{\bf 1}&\cdot\cr
              \cdot  & \cdot &\cdot&\cdot  &\cdot  &\cdot\cr
              }\right]\quad\hbox{and}\quad
Y=\left[\matrix{R &{\bf 0} &\cdot&{\bf 0} &{\bf 0} &\cdot\cr
              {\bf 0} & R&\cdot&{\bf 0} &{\bf 0} &\cdot\cr
              \cdot & \cdot &\cdot&\cdot  &\cdot  &\cdot\cr
              {\bf 0} &{\bf 0} &\cdot& R &{\bf 0} &\cdot\cr
              {\bf 0} & {\bf 0}&\cdot&{\bf 0} &R &\cdot\cr
              \cdot  & \cdot &\cdot&\cdot  &\cdot  &\cdot\cr
              }\right]\,.
\eqno(6)
$$
Hence we have $G\Psi=[{\bf z}_1,{\bf z}_2,\cdot,{\bf z}_j,\cdot;\;{\bf w}_1,{\bf w}_2,\cdots,{\bf w}_k,\cdot]^T$
where
$$
{\bf z}_j=\left[\matrix{
\sum_i p_{ji}a_i\cr \sum_i p_{ji} b_i\cr \sum_l u_{jl}\gamma_l\cr \sum_l u_{jl}\delta_l\cr}\right]
\quad\hbox{and}\quad
{\bf w}_k=\left[\matrix{
\sum_i v_{ki}a_i\cr \sum_i v_{ki} b_i\cr \sum_l q_{kl}\gamma_l\cr \sum_l q_{kl}\delta_l\cr}\right]\,.
\eqno(7)
$$
On the other hand,
$Y\Psi=[R{\bf x}_1,R{\bf x}_2,\cdots,R{\bf x}_j;\;\cdots,R{\bf y}_1,R{\bf y}_2,\cdots,R{\bf y}_k,\cdots]^T$ follows
from second equation in (6), where, taking into account (gc.2) and (gc.3),
$$
R{\bf x}_j=\left[\matrix{a_j\cr 0\cr 0\cr 0}\right]\,,\quad R{\bf y}_k=\left[\matrix{0\cr 0\cr \gamma_k\cr 0}\right],
\forall j,k\,.\eqno(8)
$$
Finally, using (7) and (6),
condition (C.4) stating $G\Psi=Y\Psi$ can be explicited as
$$
(i)\,\cases{
\sum_i p_{ji}a_i=a_j\cr \sum_i p_{ji} b_i=0\cr}\,,\;
(ii)\,\cases{\sum_l u_{jl}\gamma_l=0\cr \sum_l u_{jl}\delta_l=0\cr}\,;\eqno(gc.4)
$$
$$
(iii)\,
\cases{
\sum_i v_{ki}a_i=0\cr \sum_i v_{ki} b_i=0\cr}\,,
\quad(iv)\,\cases{\sum_l q_{kl}\gamma_l=\gamma_k\cr \sum_l q_{kl}\delta_l=0}
\,.
$$
\vskip1pc\noindent
In the representation we have adopted conditions (C.1)-(C.4) lead to general constraints
also for vectors $E\Psi$ and $G\Psi$.
Indeed, from (5), (gc.2) and (C.3) we derive
$$
E\psi=T\Psi= [a_1,b_1,0,0,\,a_2,b_2,0,0,\cdots;\;0,0,0,0,\,0,0,0,0,\cdots]^T.\eqno(gc.5)
$$
Similarly, (6), ((gc.2) and (C.4) lead to
$$
G\Psi=Y\Psi=[a_1,0,0,0,\,a_2,0,0,0,\cdots;\;0,0,\gamma_1,0,\,0,0,\gamma_2,0,\cdots]^T.\eqno(gc.6)
$$
\vskip1pc
\centerline{\bf C. Searching Solutions of ($\cal P$)}
\vskip.5pc\noindent
So far we have established general constraints to be satisfied by every solution of ($\cal P$), independently of the
ranks of $L$, $K$, $A$, $B$, $C$, $D$, and therefore of the dimension of spaces ${\cal H}_1$ and ${\cal H}_2$.
Now we begin the research of {\cal concrete} solutions of (C.1)-(C.5).
We notice that if a solution $K$ of (gc.4) exists in correspondence with
a given state vector $\Psi$ satisftying (gc.3), then all (C.1)-(C.4) are automatically satisfied
provided that $K=K^\ast=K^2$.
\par
Thus, we shall proceed as follows.
First, we fix the dimension of sèpace ${\cal H}_1$, and we find the general solutions $K$ and $\Psi$ of (gc.4).
Then, we select among these solutions of (gc.4)
the solutions of the entire problem ($\cal P$), by imposing idempotence and self-adjointness to $K$,
once the solutions which violate (C.5) have been excluded.
It is wortwhile to notice that in the case of more solutions $K_1$, $K_2$, ...,
there are several properties $G_1=K_1\otimes{\bf 1}_2$, $G_2=K_2\otimes{\bf 1}_2$, ...,
detectable together with WS property $E$, and which do not commute with $E$.
However, due to (gc.6), every $G_i$ tranforms $\Psi$ into the same vector $Y\Psi$, because $Y$ depends only
on the decomposition ${\bf 1}=A+B+C+D$.
\vskip1pc
The existence of solutions turns out to depend upon the dimension of ${\cal H}_1$. For instance no solution exists
if ${\it dim}({\cal H}_1)=2$. Indeed, in this case
$\Psi=[a,b,0,0\,;\;0,0,\gamma,\delta]^T$.
Operators $L$ and $K$ are 2$\times$2 matrices
$L=\left[\matrix{1&0\cr 0&0\cr}\right]$ and $K=\left[\matrix{p&u\cr v&q\cr}\right]$.
Since complex number $u$ must be non-vanishing to satisfy (gc.1), constraint (gc.4.ii) implies
$\gamma=\delta=0$; on the other hand, (gc.4.iii) implies $a=b=0$, thus $\Psi=0$.  Solutions of (${\cal P}$) might exist
only if {\it dim}$({\cal H}_1)>2$.
\par
In sect. III we find the solutions for
{\it dim}$({\cal H}_1)=4$, showing that whenever they exist the detections of $E$ and $G$ are always correlated.
Non-correlated solutions do exist for ${\it dim}({\cal H}_1)\geq 6$, investigated in sect. IV.
We shall restrict ourselves to the case that the two slits are symmetrical: this leads to exclude odd dimension of
${\cal H}_1$ and, moreover, to assume that ${\it rank}(L)={\it rank}({\bf 1}-L)={\dim}({\cal H}_1)/2$.
\vskip1pc
\centerline{\bf III. THE CASE {\it dim}$({\cal H}_1)=4$}
\vskip.5pc\noindent
Here we
seek for solutions of (gc.4) such that the rank of $L$ and ${\bf 1}-L$ is 2, so that $i=1,2$ and $l=1,2$.
No constraint is imposed to the ranks
of $A$, $B$, $C$ and $D$, and hence to the dimension of ${\cal H}_2$. If ${\it dim}({\cal H}_1)=4$, then
$\Psi=[{\bf x}_1,{\bf x}_2;\;{\bf y}_1,{\bf y}_2]^T$, so that $P$, $U$, $V$ and $Q$ are
$2\times 2$ matrices. Then (gc.4.ii) implies
$u_{11}{\bf y}_1+u_{12}{\bf y}_2=0$ and $u_{21}{\bf y}_1+u_{22}{\bf y}_2=0$.
Therefore, since $U\neq{\bf 0}$ by (gc.1),
the vectors ${\bf y}_1$ and ${\bf y}_2$ must be linearly dependent. Let us suppose that
$\gamma_2=\lambda\gamma_1$ and $\delta_2=\lambda\delta_1$, with $\lambda\neq 0$. By using these relations in
(gc.4.iv) we get
$$
\cases{q_{11}\gamma_1+q_{12}\gamma_2=\gamma_1=(q_{11}+\lambda q_{12})\gamma_1\cr
q_{21}\gamma_1+q_{22}\gamma_2=\gamma_2=(q_{21}+\lambda q_{22})\gamma_1\cr
q_{11}\delta_1+q_{12}\delta_2=0=(q_{11}+\lambda q_{12})\delta_1\cr
q_{21}\delta_1+q_{22}\delta_2=0=(q_{21}+\lambda q_{22})\delta_1\,.\cr}
$$
If $\delta_1\neq 0$ then $(q_{11}+\lambda q_{12})=(q_{21}+\lambda q_{22})=0$, which implies $\gamma_1=\gamma_2=0$.
On the other hand, if $\delta_1=0$ then $\delta_2=\lambda\delta_1=0$, while $\gamma_1$, $\gamma_2$ can be non-vaninshing
with $\gamma_2=\lambda\gamma_1$.
Similarly, from (gc.4.iii) it follows that ${\bf x}_1$ and ${\bf x}_2$ are linearly dependent, say
${\bf x}_2=\mu{\bf x}_1$, and we can establish that if $b_1\neq 0$ then $a_1=a_2=0$;
if $b_1=0$ then $b_2=0$ and $a_2=\mu a_1$. More explicitly,
\vskip1pc\noindent
a) $\Psi=[a_1,0,0,0,\,\mu a_1,0,0,0;\;0,0,\gamma_1,0,\,0,0,\lambda\gamma_1,0]^T$, if $b_1=0$ and $\delta_1=0$;
\vskip1pc\noindent
b) $\Psi=[a_1,0,0,0,\,\mu a_1,0,0,0;\;0,0,0,\delta_1,\,0,0,0,\lambda\delta_1]^T$, if $b_1=0$ and $\delta_1\neq 0$;
\vskip1pc\noindent
c) $\Psi=[0,b_1,0,0,\,0,\mu b_1,0,0;\;0,0,\gamma_1,0,\,0,0,\lambda\gamma_1,0]^T$, if $b_1\neq 0$ and $\delta_1= 0$;
\vskip1pc\noindent
d) $\Psi=[0,b_1,0,0,\,0,\mu b_1,0,0;\;0,0,0,\delta_1,\,0,0,0,\lambda\delta_1]^T$, if $b_1\neq 0$ and $\delta_1\neq 0$.
\vskip1pc\noindent
General constraints (gc.5) and (gc.6) imply that
case (a) and (d) violate (C.5) because they respectively yield $G\Psi=\Psi$ and  $G\Psi=0$.
Therefore, if a property $G$ can be detected in such a way that (C.1)-(C.5)
hold, the state vector must have one of the forms (b)-(c).
\par
Before entering the question of the existence of such a property $G$, we can draw the following conclusion.
Let us suppose that $G$ exists, so that a detection of
such a property can be carried out together with WS-detection, i.e. without erasing WS knowledge, and
without affecting the point of impact on the final screen. If case (b) for $\Psi$ is realized then, according
to (gc.5) and (gc.6)
we have
$$
T\Psi= [a_1,0,0,0,\,\mu a_1,0,0,0,\cdots;\;0,0,0,0,\,0,0,0,0,\cdots]^T=Y\Psi.\eqno(9)
$$
As a consequence, both conditional probabilities $p(T\mid
Y)=\langle\Psi\mid TY\Psi\rangle/\langle\Psi\mid Y\Psi\rangle$ and
$p(Y\mid T)=\langle\Psi\mid YT\Psi\rangle/\langle\Psi\mid
T\Psi\rangle$ are equal to 1. Then we can conclude that
\begin{description}
\item[\qquad]
{\sl property $G$ is detected by $Y$ on a particle (i.e. the outcome for $Y$ is 1) if and only if $T$ detects
the passage of that particle through slit 1.}
\end{description}
Thus,
in the present case (${\cal H}_1\equiv{\bf C}^4$), for each solution of problem (${\cal P}$) there is a direct
correlation between the detection of $G$ and the passage through slit 1, so that detector $Y$ does not give rise to a
sorting different from that carried out by WS detector $T$.
\vskip1pc
Now we face the problem of singling out solutions $K$ of (gc.4) in the case of interest (b), where
$\delta_1\neq 0$ and $b_1=0$. In sub-section A we find the solutions corresponding to $\mu=0$ (or $\lambda=0$), whereas
solutions for $\mu\neq 0\neq\lambda$ are singled out in sub-section B.
\par
Case (c) can be treated in a similar way, giving rise to quite symmetrical results.
\vskip1pc
\centerline{\bf A. The case $\mu=0$ or $\lambda=0$}
\vskip.5pc\noindent
If $\mu=0$ then $\Psi=[a_1,0,0,0,\,0,0,0,0;\;0,0,0,\delta_1,\,0,0,0,\lambda\delta_1]^T$.
Since $a_1=0$ implies $T\Psi=0$, and $\delta_1=0$ implies $T\Psi=\Psi$, which violate (C.5), we have to
consider only the case that both $a_1$ and $\delta_1$ are non-vanishing.
Therefore,
\par\noindent
-- (gc.4.i) implies $p_{11}=1$, $p_{21}=0=p_{12}$,
\par\noindent
-- (gc.4.iii) implies $v_{11}=v_{21}=0$, and $u_{11}=u_{12}=0$ by self-adjointness,
\par\noindent
--
(gc.4.ii) implies $u_{21}=-\lambda u_{22}$
\par\noindent
--
(gc.4.iv) implies $q_{21}=-\lambda q_{22}$, $q_{11}=-\lambda q_{12}=-\lambda \overline{q_{21}}=\vert\lambda\vert^2 q_{22}$.
\par\noindent
Then matrix $K$ must have the form
$$
K=\left[
\matrix{1&0&0&0\cr
0&p&-\lambda u& u\cr
0&-\overline\lambda\overline u&\vert\lambda\vert^2 q&-\overline\lambda q \cr
0&\overline u&-\lambda q&q\cr}\right],
\eqno(10)
$$
where $q\equiv q_{22}$ and $u\equiv u_{22}$. Now we find the solutions such that
{\it rank}$(K)=1+p+\vert\lambda\vert^2+q=2$.
Imposing idempotence to elements $p$ and $q$ we
have that given any $\lambda$, matrix $K$ in (10) is a solution if and only if
$u=e^{i\theta}\sqrt{p-p^2\over 1+\vert\lambda\vert^2}$, $q={1\over 1+\vert\lambda\vert^2}$ and
$0<p<1$.
\par
The case $\lambda=0$ can be treated along the same logical lines, and lead to quite symmetrical results.
\vskip1pc
\centerline{\bf B. The case $\mu\neq 0\neq\lambda$}
Now we consider the remaining case $\mu\neq 0\neq\lambda$. Self-adjointness together with (gc.4) leads to
$$
K=\left[
\matrix{1-\vert\mu\vert^2(1-p)&\bar\mu(1-p)&-\lambda u& u\cr \mu(1-p)& p &{\lambda\over\bar\mu}u & -{1\over\bar\mu}u\cr
-\bar\lambda\bar u& {\bar\lambda\over\mu}\bar u&\vert\lambda\vert^2 q&-{\bar\lambda}q\cr
\bar u&-{1\over\mu}\bar u&-\lambda q&q\cr}\right].
\eqno(11)
$$
Now, matrix $K$ in (11) is a solution if
numbers $p$, $q$, $u\neq 0$ can be chosen in such a way that $K$ turns out to be idempotent.
\par
The set of solutions is not empty.
We show the solutions for
the case $\lambda=\mu=1$ and {\it rank}$(K)=2$, so that (11) becomes
$
K=\left[\matrix{p&1-p&-u&u\cr 1-p&p&u&-u\cr -\bar u&\bar u& q&-q\cr \bar u&-\bar u&-q&q\cr}\right].
$
Idempotence imposes $0<q<1/2$, $p=1-q$ and $u=e^{i\theta}\sqrt{({1\over 2}-q)q}$, where $\theta$ is any real number.
Therefore, for every $q\in(0,1/2)$ and every $\theta\in{\bf R}$ we have a solution
$$
K=\left[\matrix{1-q&q&-e^{i\theta}\sqrt{({1\over 2}-q)q}&e^{i\theta}\sqrt{({1\over 2}-q)q}\cr
q&1-q&e^{i\theta}\sqrt{({1\over 2}-q)q}&-e^{i\theta}\sqrt{({1\over 2}-q)q}\cr
-e^{-i\theta}\sqrt{({1\over 2}-q)q}&e^{-i\theta}\sqrt{({1\over 2}-q)q}& q&-q\cr
e^{-i\theta}\sqrt{({1\over 2}-q)q}&-e^{-i\theta}\sqrt{({1\over 2}-q)q}&-q&q\cr}\right].
\eqno(12)
$$
In [17] is presented the particular solution corresponding to $\theta=0$ and $q=1/4$.
\vskip1pc
\centerline{\bf IV. THE CASE ${\it dim}({\cal H}_1)=6$.}
\vskip.5pc\noindent
In the case {\it dim}$({\cal H}_1)=4$ investigated in sect. III we saw that WS-passage can be detected
by a non-disturbing WS detector $T$ together with the detection of a property $G=K\otimes{\bf 1}$
which does not commute with $E$. But the two detections, whenever are possible, are directly correlated,
so that the detection of $G$ by $Y$ sorts exactly the particles, and only those, detected by $T$.
In this section we want to answer the question whether,
by allowing the dimension of ${\cal H}_1$ to be greater than 4,
properties $G$ can be singled out
which are detectable together with $E$ but which are not correlated with $E$.
\par
We assume that {\it rank}$(L)=$\,{\it rank}$({\bf 1}_1-L)=3$,
so that $\Psi=[{\bf x}_1,{\bf x}_2,{\bf x}_3;\;{\bf y}_1,{\bf y}_2,{\bf y}_3]^T$ and
in (gc.4) indexes $i,j,k,l$ take values in $\{1,2,3\}$. Since $U\neq{\bf 0}$, general constraint
(gc.4.ii) implies that the three vectors
${\bf y}_1,{\bf y}_2,{\bf y}_3$ are linearly dependent, so that two complex numbers $\lambda_{31},\lambda_{32}$
must exist such that
$$
\cases{\delta_3=\lambda_{31}\delta_1+\lambda_{32}\delta_2\cr
\gamma_3=\lambda_{31}\gamma_1+\lambda_{32}\gamma_2\cr}.
\eqno(13)
$$
Using these equations in (gc.4.iv) we get
$$
\cases{(q_{k1}+\lambda_{31}q_{k3})\delta_1+(q_{k2}+\lambda_{32}q_{k3})\delta_2=0\cr
\gamma_k=(q_{k1}+\lambda_{31}q_{k3})\gamma_1+(q_{k2}+\lambda_{32}q_{k3})\gamma_2}, \;k=1,2,3.
\eqno(14)
$$
If vectors $\delta_1$, $\delta_2$ are linearly independent, then first equation in
(14) implies $(q_{k1}+\lambda_{31}q_{k3})=(q_{k2}+\lambda_{32}q_{k3})=0$, so that second equation in (14)
yields $\gamma_k=0$ for all $k$. Hence
$$
\delta_1,\delta_2\;\hbox{linearly independent}\quad\Rightarrow\quad{\bf y}_k=[0,0,0,\delta_k]^T,\;\forall k.
\eqno(15)
$$
In a similar way we can prove that
$$
b_1,b_2\;\hbox{linearly independent}\quad\Rightarrow\quad{\bf x}_k=[0,b_k,0,0]^T,\;\forall k.
\eqno(16)
$$
Now we draw the consequences of (15) and (16) relative to our
problem (${\cal P}$). Given a state vector $\Psi$ satisfying
general constraint (gc.3), a possibility is that $\delta_1$,
$\delta_2$ are linearly independent and also $b_1$, $b_2$ are
linearly independent. In this case (15), (16) imply ${\bf
x}_k=[0,b_k,0,0]^T$ and ${\bf y}_k=[0,0,0,\delta_k]^T$. If a
solution of (${\cal P}$) exist, then $G\Psi=Y\Psi=0$ would follow
from (gc.6). Therefore, meaningful solutions of problem (${\cal
P}$) cannot be found in this case because (C.5) is violated. If we
consider all possible cases, then we obtain the following
implications. \vskip1pc\noindent {a)} $\;\;\delta_1$, $\delta_2$
linearly independent and $b_1$, $b_2$ linearly independent imply
$G\Psi=0$. \vskip.5pc\noindent {b)} $\;\;\delta_1$, $\delta_2$
linearly independent and $b_1$, $b_2$ linearly dependent imply
\par
${\bf x}_k=[a_k,b_k,0,0]^T$ and ${\bf y}_k=[0,0,0,\delta_k]^T$.
\vskip.5pc\noindent {c)} $\;\;\delta_1$, $\delta_2$ linearly
dependent and $b_1$, $b_2$ linearly independent imply
\par
${\bf x}_k=[0,b_k,0,0]^T$ and ${\bf
y}_k=[0,0,\gamma_k,\delta_k]^T$. \vskip.5pc\noindent {d)}
$\;\;\delta_1$, $\delta_2$ linearly dependent and $b_1$, $b_2$
linearly dependent imply
\par
${\bf x}_k=[a_k,b_k,0,0]^T$ and ${\bf y}_k=[0,0,\gamma_k,\delta_k]^T$.
\vskip.5pc\noindent
We shall search solutions for cases (b), (c) and (d), since in case (a) meaningful solutions cannot exist.
\vskip1pc
\centerline{\bf A. Cases (b) and (c).}
\vskip.5pc\noindent
According to (b), we can state that if a solution of (${\cal P}$) exists such that
$\delta_1$, $\delta_2$ are linearly independent and $b_1$, $b_2$ are linearly dependent, then
(gc.5), (gc.6) imply
$YT\Psi=Y\Psi$ holds, which is equivalent to say that conditional probablity
$p(T\mid Y)=\langle\Psi\mid TY\Psi\rangle / \langle\Psi\mid Y\Psi\rangle$ is equal to 1;
this means that each time a particle is sorted by $T$, then it is certainly sorted by $Y$.
\par
In case (c) $TY\Psi=T\Psi$ holds, so that each time a particle is sorted by $Y$, then it is certainly sorted by $T$.
Therefore, for all eventual solutions corresponding to cases (b) and (c), property $G$ must be correlated with WS property $E$.
The only case which can lead to solution without correlation is case (d).
\vskip1pc
\centerline{\bf B. Case (d)}
\vskip.5pc\noindent
In case (d) we may suppose that
$b_2=\mu b_1$ and $\delta_2=\lambda \delta_1$.
If $b_k\neq 0$ and $\gamma_k\neq 0$ for at least a $k$, then $T\Psi\neq TY\Psi\neq Y\Psi$.
This means that no correlation exists between the detections of properties $E$ and $G$ carried out by means of $T$ and $Y$ respectively.
\par
We shall see that concrete solutions $K$ of (C.1)-(C.4) exist in this case. Our task is more simple if we search solutions
corresponding to {\sl particular} state vectors $\Psi$ satisfying (d), (which implies (gc.3)). Hence we search solutions
corresponding to vector state $\Psi$ such that
$$\cases{
a_1=a_2=0,\, a_3\neq 0, \;b_2=\mu b_1\neq 0,\,b_3=0,\cr  
\gamma_1=\gamma_2=0,\,\gamma_3\neq 0,\;\delta_2=\lambda\delta_1\neq 0,\,\delta_3=0.}
\leqno(17)
$$
Then (gc.4.i) and (gc.4.iv) respectively imply
$p_{13}=p_{23}=0$, $p_{33}=1$ and $q_{13}=q_{23}=0$, $q_{33}=1$, so that matrices $P$ and $Q$
in $K=\left[\matrix{P&U\cr V&Q\cr}\right]$
have the form
$$
P=\left[\matrix{p_{11}&p_{12}&0\cr p_{21}&p_{22}&0\cr 0&0&1}\right]\quad\hbox{and}\quad
Q=\left[\matrix{q_{11}&q_{12}&0\cr q_{21}&q_{22}&0\cr 0&0&1}\right].
\eqno(18)
$$
Similarly, (gc4.iii) implies $v_{13}=v_{23}=v_{33}=0$ and hence, by the self-adjointness of $K$,
$u_{31}=u_{32}=u_{33}=0$. On the other hand, the first equation in (gc.4.ii) and (17) imply
$u_{13}=u_{23}=0$ and hence $v_{31}=v_{33}=0$. Therefore, matrices $V$ and $U$ have the form
$$
V=\left[\matrix{v_{11}&v_{12}&0\cr v_{21}&v_{22}&0\cr 0&0&0\cr}\right]\quad\hbox{and}\quad
U=\left[\matrix{u_{11}&u_{12}&0\cr u_{21}&u_{22}&0\cr 0&0&0\cr}\right].
\eqno(19)
$$
Taking into account (17), (18), (19), (gc.4) become
$$
(i)\;\cases{p_{11}+\mu p_{12}=0\cr p_{21}+\mu p_{22}=0},\quad
(ii)\;\cases{u_{11}+\lambda u_{12}=0\cr u_{21}+\lambda u_{22}=0},
\eqno(20)
$$
$$
(iii)\;
\cases{v_{11}+\mu v_{12}=0\cr v_{21}+\mu v_{22}=0},\quad (iv)\;
\cases{q_{11}+\lambda q_{12}=0\cr q_{21}+\lambda q_{22}=0}.
$$
The self-adjointness of $K$, together with (20),
yields
$$
K=
\left[\matrix{p &-{p/\mu}&0&u&-{u/\lambda}&0\cr
-{p/\bar\mu}&{p/\vert\mu\vert^2}&0&-{u/\bar\lambda}& {u/\vert\lambda\vert^2}&0\cr
0&0&1&0&0&0\cr
\bar u&-{\bar u/\lambda}&0&q&-{q/\lambda}&0\cr
-{\bar u/\bar\lambda}&{\bar u/\vert\lambda\vert^2}&0&-{q/\bar\lambda}&{q/\vert\lambda\vert^2}&0\cr
0&0&0&0&0&1\cr}\right],
\eqno(21)
$$
where we have put $p=p_{11}$, $u=u_{11}$, $v=v_{11}$, $q=q_{11}$.
By imposing idempotence we find that in correspondence with $\lambda=\mu=1$ there is a solution
$$
K=\left[
\matrix{p&-p&0&e^{i\theta}\sqrt{p({1\over 2}-p)}&-e^{i\theta}\sqrt{p({1\over 2}-p)}&0\cr
-p&p&0&-e^{i\theta}\sqrt{p({1\over 2}-p)}&e^{i\theta}\sqrt{p({1\over 2}-p)}&0\cr
0&0&1&0&0&0\cr
e^{-i\theta}\sqrt{p({1\over 2}-p)}&-e^{-i\theta}\sqrt{p({1\over 2}-p)}&0&({1\over 2}-p)&-({1\over 2}-p) & 0\cr
-e^{-i\theta}\sqrt{p({1\over 2}-p)}&e^{-i\theta}\sqrt{p({1\over 2}-p)}&0&-({1\over 2}-p)&({1\over 2}-p)&0\cr
0&0&0&0&0&1 }\right]
\eqno (22)
$$
such that {\it rank}$(K)=3$, for any $p$ such that $0<p<1/2$ and
any $\theta\in{\bf R}$. For instance, the following solution of
($\cal P$)
$$
\begin{array}{rcl}
\Psi&=&[0,b_1,0,0,\; 0,b_1,0,0,\;a_3,0,0,0; 0,0,0,\delta_1,\;0,0,0,\delta_1,\; 0,0,\gamma_3,0]^T \\
K&=&
\left[\matrix{{1\over 4}&-{1\over 4}&0&{1\over 4}&-{1\over 4}&0\cr -{1\over 4}&{1\over 4}&0&-{1\over 4}&{1\over 4}&0\cr
0&0&1&0&0&0\cr
{1\over 4}&-{1\over 4}&0&{1\over 4}&-{1\over 4}&0\cr -{1\over 4}&{1\over 4}&0&-{1\over 4}&{1\over 4}&0\cr
0&0&0&0&0&1\cr}\right].\\
\end{array}
\eqno(23)
$$
is obtained in correspondence with the particularly simple choice $\theta=0$ and
$p=1/4$.
\vskip1pc
\centerline{\bf V. NO INTERFERENCE THEOREM}
\vskip.5pc\noindent
In this section we investigate the relationships between non-disturbing detection and
presence or absence of interference fringes.
\par
The appearence of interference in double-slit experiment is a typical quantum phenomenon, which stresses the departure
of Quantum Physics from a pre-quantum conception of natural phenomena.
Before studying what happens to interference when non-disturbing detectors exist (subsect. B), we theoretically explain
the emergence of intereference, singling out assumption ($\cal C$) below as the point of departure from classical description.
\vskip1pc
\centerline{\bf A. Interference excludes WS property} 
\vskip.5pc
At time $t_1$, i.e. when the particle crosses the support of the slits, quantum theory prescribes
well defined probabilities
$\pi(1)=\langle\Psi\mid E\Psi\rangle$ and
$\pi(2)=\langle\Psi\mid E'\Psi\rangle$ for the passing through respectively
slit 1 or 2. Moreover, since these two events are represented by mutually orthogonal projections, $E\perp E'$,
we may state that at time $t_1$ the particle can be observed to pass either through
slit 1 or through slit 2.
A pre-quantum attitude would lead to infer the following assumption
\vskip.5pc\noindent
{($\cal C$)}\quad
{\sl Each particle, considered at a time $t>t_1$, passed at time $t_1$ either through slit 1 or through slit 2,
with respective probabilities $\pi(1)$ and $\pi(2)$.}
\vskip.5pc\noindent
But some consequence of ($\cal C$), as (25) below, contradict the predictions of a
quantum theoretical treatment.
\par
If ($\cal C$) held together with quantum theory, for every projection operator $F$ there should
be a conditional probability $p(F\mid E)$ carrying the natural
properties of conditional probability, i.e.
\vskip.5pc\noindent
{i)}\quad
if $F=\sum_i F_i$, where $F_i\perp F_j$ for $i\neq j$, then
$p(F\mid
E)=\sum_ip(F_i\mid E)$,
\hfill{(additivity)}
\vskip.5pc\noindent
{ii)}\quad
$F\leq E$ implies
$p(F\mid E)=\langle\Psi\mid F\Psi\rangle$.
\hfill{(consistency with quantum probabilities)}
\vskip.5pc\noindent A
theorem proved by Cassinelli and Zangh\`\i\/ [18] states that if
$p(F\mid E)$ satisfies (i) and (ii) then
$$
p(F\mid E)={\langle \Psi\mid EFE\Psi\rangle\over\langle\Psi\mid E\Psi\rangle}=
\langle {\Psi_1\over\Vert\Psi_1\Vert}\mid F{\Psi_1\over\Vert\Psi_1\Vert}\rangle,
\eqno(24.i)
$$
where $\Psi_1=E\Psi$.
The same argument for $E'$ yields
$$
p(F\mid E')={\langle \Psi\mid E'FE'\Psi\rangle\over\langle\Psi\mid E'\Psi\rangle}=
\langle {\Psi_2\over\Vert\Psi_2\Vert}\mid F{\Psi_2\over\Vert\Psi_2\Vert}\rangle,\quad  \Psi_2=E'\Psi.
\eqno(24.ii)
$$
Therefore, the particles coming from slit $i$ are represented by the state vector
${\Psi_i\over\Vert\Psi_i\Vert}$.
For $F=F(\Delta)$, from (e.40) it follows that
$\langle\Psi_i\mid F(\Delta)\Psi_i\rangle=\pi(i)\langle {\Psi_i\over\Vert\Psi_i\Vert}\mid F(\Delta)
{\Psi_i\over\Vert\Psi_i\Vert}\rangle$ is the probability that the particle hits $\Delta$ passing through slit $i$;
then assumption ($\cal C$) implies that the probability
that the particle hits $\Delta$
should be
$$
p_{\cal C}(F(\Delta))=\langle\Psi_1\mid
F(\Delta)\Psi_1\rangle+\langle\Psi_2\mid F(\Delta)\Psi_2\rangle.
\eqno(25)
$$
\par
On the other hand the correct quantum theoretical prediction for this probability is
$$
\begin{array}{rcl}
p(F(\Delta))&=&\langle\Psi\mid F(\Delta)\Psi\rangle=\langle\Psi_1\mid F(\Delta)\Psi_1\rangle
+\langle\Psi_2\mid F(\Delta)\Psi_2\rangle+\\
&&+2\,{\it Re}(\langle\Psi_1\mid F(\Delta)\Psi_2\rangle)\\
\end{array}
\eqno(26)
$$
We see that if the {\sl interference term}
$2\,{\it Re}(\langle\Psi_1\mid
F(\Delta)\Psi_2\rangle)$ is different from 0, a contradiction occurs between the purely
quantum theoretical prediction (26) and prediction (25) implied by additional assumption ($\cal C$).
Therefore, according to standard intepretation of quantum theory,
interference forbids to assign WS property to each particle which hits the final screen; in other words, the presence of interference fringes,
experimentally verified in agreement with quantum predictions also for mesoscopic
systems, as {\sl fullerene} molecules [19],
{\sl erases} WS property (fig. 2).
\begin{figure}[htb!]
\centerline{\epsfig{figure=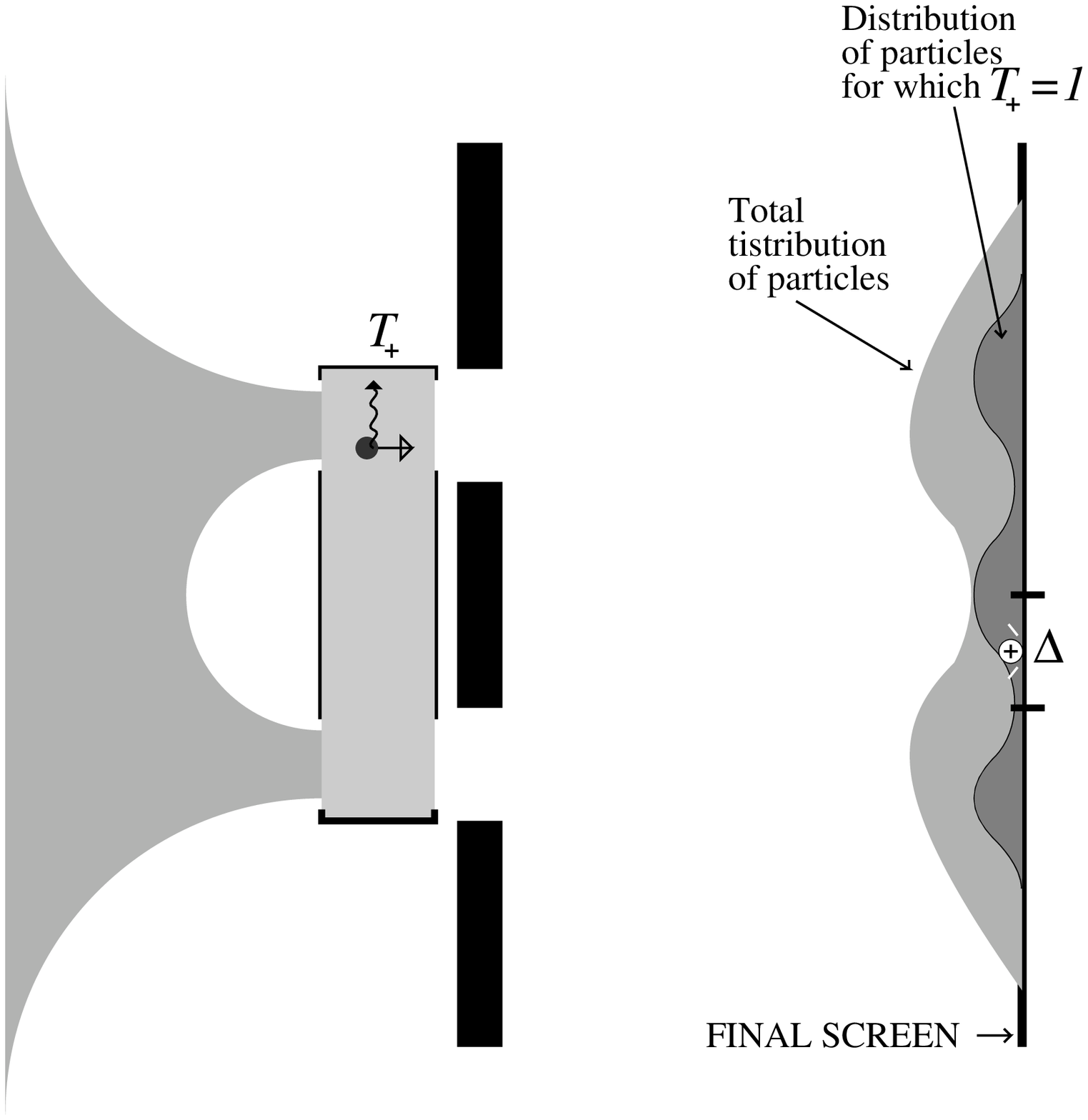,height=8cm}} \caption[]{
Erasure} \label{fig2}
\end{figure}
\vskip1pc
\centerline{\bf B. Interference and erasure}
\vskip.5pc\noindent
Now we proceed in investigating interference in presence of non-disturbing detectors.
If a WS non-disturbing detector exists with respect to the state vector $\Psi$ describing the particle,
it is possible to ascertain which slit each particle hitting $\Delta$ comes from. Therefore assumption ($\cal C$)
holds in such a case, and no interference can take place. This argument admits a direct, simple formal proof:
$\langle\Psi_1\mid F(\Delta)\Psi_2\rangle=\langle E\Psi\mid F(\Delta) E'\Psi\rangle=
\langle T\Psi\mid F(\Delta) T'\Psi\rangle=\langle \Psi\mid F(\Delta)T T'\Psi\rangle=0$, because $[T,F(\Delta)]=0$.
Therefore, the absence of interference shown by ESW in their thought experiment when WS detector is used,
does not depend on the particular set-up,
but it is a general feature implied by the existence of a non-disturbing WS detector.
\vskip1pc
Now we want to see what happens about interference if only the particles detected by a detector of the kind
$Z={\bf 1}\otimes Q$ are considered.
Their distibution is ruled over by the probability $p(F(\Delta)\land Z)$
that a particle hits $\Delta$ and it is also detected by $Z$.
According to quantum theory, since $[Z,F(\Delta)]={\bf 0}$
the joint event
$F(\Delta)\land Z$ is represented by projection
$Z F(\Delta)$. Then we have
$$
\begin{array}{rcl}
p(F(\Delta)\land Z)&=&\langle\Psi\mid ZF(\Delta)\Psi\rangle=\langle\Psi_1+\Psi_2\mid ZF(\Delta)(\Psi_1+\Psi_2)\rangle\\
&=&\langle\Psi_1\mid Z F(\Delta)\Psi_1\rangle
+\langle\Psi_2\mid Z F(\Delta)\Psi_2\rangle+\\
&&+2\,{\it Re}(\langle\Psi_1\mid ZF(\Delta)\Psi_2\rangle.\\
\end{array}
$$
We see that
interference is present if and only if the last term is not vanishing.
It is interesting to notice that the absence of interference for all particles, i.e. without performing selection by some
detector $Z$, does not imply the absence of interference for the selected particles. The thought experiment
proposed by ESW, outlined here in examples 1 and 2, provides a clear demonstration of this phenomenon.
Indeed, the existence of the non disturbing WS detector $\;T={\bf 1}\otimes\vert 1\rangle\langle 1\vert$
implies that the interference term $2\,{\it Re}(\langle\Psi_1\mid F(\Delta)\Psi_2\rangle$ is 0.
But, if we select on the final screen those particles detected by the non-disturbing detector
$T_+={\bf 1}\otimes\vert+\rangle\langle+\vert$, we have
$p(F(\Delta)\land T_+)=\langle\Psi_1\mid T_+ F(\Delta)\Psi_1\rangle
+\langle\Psi_2\mid T_+ F(\Delta)\Psi_2\rangle+(1/2){\it Re}(\langle\psi_1\mid J\psi_2\rangle$; owing to
${\it Re}(\langle\psi_1\mid J\psi_2\rangle\neq 0$ (see example 1), we conlude that
interference reappears, as shown in fig. 2.
\par
Now we suppose that the selection on the final screen is performed
by means of detector $Y$ taken in a solution of problem (${\cal
P}$). In such a case interference cannot reappear, indeed,
$\langle\Psi_1\mid YF(\Delta)\Psi_2\rangle=\langle E\Psi\mid
YF(\Delta) E'\Psi\rangle= \langle T\Psi\mid YF(\Delta)
T'\Psi\rangle=\langle \Psi\mid YF(\Delta)T T'\Psi\rangle=0$,
because $[F(\Delta),T]=[Y,T]={\bf 0}$. We formulate this statement
as a theorem.
\vskip.5pc\noindent
{\bf No interference Theorem.}
{\sl Let $T$ be WS non-disturbing
detector with respect to $\Psi$. If $Y$ is a solution of ($\cal
P$), then the particles detected by $Y$ do not give rise to
interference.} 
\vskip1pc\noindent
\centerline{\bf VI. CONCLUDING REMARKS.}
\vskip.5pc\noindent
Our treatment has been entirely carried out on a theoretical ground. A complete scientific assessment
of the results requires the possibility of experiments for confirming or rejecting the theoretical predictions. The problem of designing concretely realizable experiments goes beyond the matter covered by the present work. However,
we shall describe an ideal apparatus which exploits the physical principles used by ESW to devise their
thought experiment.
Our experimental setup corresponds to the particular solution (23), where
$$
\begin{array}{rcl}
\Psi&=&[0,b,0,0,\; 0,b,0,0,\;a,0,0,0; 0,0,0,\delta,\;0,0,0,\delta,\; 0,0,\gamma,0]^T \\
K&=&
\left[\matrix{{1\over 4}&-{1\over 4}&0&{1\over 4}&-{1\over 4}&0\cr -{1\over 4}&{1\over 4}&0&-{1\over 4}&{1\over 4}&0\cr
0&0&1&0&0&0\cr
{1\over 4}&-{1\over 4}&0&{1\over 4}&-{1\over 4}&0\cr -{1\over 4}&{1\over 4}&0&-{1\over 4}&{1\over 4}&0\cr
0&0&0&0&0&1\cr}\right].\\
\end{array}
\eqno(27)
$$
Let us think each of the two slits as decomposed into 3 regions, up ({\bf u}), centre ({\bf c}), down ({\bf d}).
The system consists of an atom in a long lived excited state as in example 1. The position of its centre-of-mass is described
in space ${\cal H}_1$. The further degrees of freedom, described by means of ${\cal H}_2$, concern with
four (rather than 2 as in example 1) micromaser cavities $\hat A$, $\hat B$, $\hat C$, $\hat D$, placed as in fig. 3.
By $A$ we denote the projection operator of ${\cal H}_2$ representing the event
``a photon is revealed in cavity $\hat A$''. In such a way, we can define four projections $A$, $B$, $C$, $D$
associated to all cavities $\hat A$, $\hat B$, $\hat C$, $\hat D$, and we shall denote their respective
eigenvectors relative to eigenvalue 1 by
$\vert a\rangle$,  $\vert b\rangle$, $\vert \gamma\rangle$ and $\vert \delta\rangle$.
As a consequence, the state vector of the entire system is
$$
\Psi={1\over\sqrt 6}\left\{(\psi^{\bf u}_1+\psi^{\bf c}_1)\vert a\rangle+\psi^{\bf d}_1\vert b\rangle
+
(\psi^{\bf u}_2+\psi^{\bf c}_2)\vert \delta\rangle+\psi^{\bf d}_2\vert \gamma\rangle\right\},
\eqno(28)
$$
where $\psi_i^{\bf u}$, $\psi_i^{\bf c}$ and $\psi_i^{\bf d}$ are normalized state vectors of ${\cal H}_1$
respectively localized in region {\bf u}, {\bf c} and {\bf d} of slit $i$.
\begin{figure}[htb!]
\centerline{\epsfig{figure=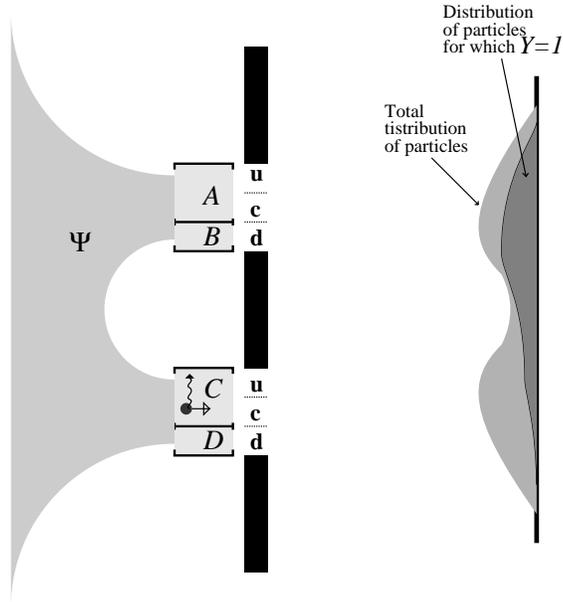,height=8cm}} \caption[]{
Ideal apparatus for detecting both $E$ and $G$} \label{fig2}
\end{figure}
\par
Within the representation adopted in the present work, the state vector in (28) concides with that in (27).
Therefore, according to the results obtained in section IV, with respect to such a state vector there are both
a non-disturbing WS detector $T={\bf 1}\otimes(A+B)$ and a non-disturbing detector
$Y={\bf 1}\otimes (A+C)$ of a property $G=K\otimes{\bf 1}$, incompatible with WS property $E$, where $K$
is the projection in (27).
\par
Therefore,
from the knowledge
of which cavity the photon is revealed in, we can infer both which slit the atom comes from and whether
it possesses either property $G$ or $G'$, according to the following scheme
$$
\matrix{
{\rm cavity}& & & & & \cr
\hat A&\Rightarrow&{\rm slit}& 1&{\rm and}& G\cr
\hat B&\Rightarrow&{\rm slit}& 1&{\rm and}& G'\cr
\hat C&\Rightarrow&{\rm slit}& 2&{\rm and}& G\cr
\hat D&\Rightarrow&{\rm slit}& 2&{\rm and}& G'\cr
}
\eqno(29)
$$
Thus, the detection of property $G$ is attained without erasing
WS knowledge.
\vskip.5pc
We have to stress the ideal character of
the experiment just described. It exploits the physical
properties of micromaser cavities, while the actually performed experiments,
for realizing WS non-disturbing detection and erasure,
make use of photon pairs produced in spontaneous parametric down
conversion [20] or, more recently, single photons [13].
\par
At the best of our knowledge, a real experiment for
simultaneous detection of WS
passage and of an incompatible property $G$ is yet to be perfomed.
A contribution to increase the possibility of a concrete realization of such an experiment,
may be to provide a richer
set of solutions of problem ($\cal P$). We notice that the set of
solutions singled out in the present work is not exhaustive. Thus,
a theoretical research for finding further solutions should be worth also
in the perspective of designing
a real experiment.
\vfill\eject
\centerline{\bf REFERENCES}
\noindent
\begin{description}
  \item[{\rm[1]}] Institut International de Physique Solvay, Rapport et discussions du 5$^e$
  Conseil, Paris 1928.
  \item[{\rm[2]}] N. Bohr, in {\sl Quantum Theory of measurement}, J.A.
  Wheeler and W.H.Zurek, eds., p.9,
  Princeton University Press, Prineton, New Jersey, 1983.
  \item[{\rm[3]}] R.P. Feynman, A.R. Hibbs, {\sl Quantum mechanics and
  path integrals}, Mc Graw-Hill inc., New York 1965.
  \item[{\rm[4]}] M.O. Scully, B.-G. Englert, H. Walther, Nature
  (London), {\bf 351} (1991) 111.
  \item[{\rm[5]}] M.O. Scully, H. Walther, Phys.Rev.A, {\bf 39} (1989) 5229.
  \item[{\rm[6]}] E.P. Storey, S.M. Tan, M.J. Collett, D.F. Walls,
  Nature (London), {\bf 367} (1994) 626.
  \item[{\rm[7]}] B.-G. Englert, M.O. Scully, H. Walther,
  Nature (London) {\bf  375} (1995) 367.
  \item[{\rm[8]}] E.P. Storey, S.M. Tan, M.J. Collett, D.F. Walls, Nature (London),
  {\bf 375} (1995) 368.
  \item[{\rm[9]}] H.M. Wiseman, F.E. Harrison, Nature (London), {\bf 377} (1995) 584.
  \item[{\rm[10]}] U. Mohrhoff, Am.J.Phys., {\bf 64} (1996) 1468.
  \item[{\rm[11]}] B.-G. Englert, M.O. Scully, H. Walther, Am.J.Phys., {\bf 67} (1999) 325.
  \item[{\rm[12]}] G. Jaeger, A. Shimony, L. Vaidman, Phys.Review A, {\bf 51} (1995) 54.
  \item[{\rm[13]}] P.D.D. Schwindt, P.G. Kwiat, B.-G. Englert, Phys.Review A, {\bf 60} (1999) 4285.
  \item[{\rm[14]}] B.-G. Englert, J. Schwinger, M.O. Scully, in {\sl New frontiers in
  quantum electrodynamics and quantum optics}, p. 507, A.O. Barut ed., Plenum, New York 1990.
  \item[{\rm[15]}] G. Nistic\`o, M.C. Romania, J.Math.Phys., {\bf 35} (1994) 4534.
  \item[{\rm[16]}] J.M. Jauch, {\sl Foundations of quantum mechanics},
  Addison-Wesley Pub.Co., Reading, Massachusets 1968.
  \item[{\rm[17]}] G. Nistic\`o, A. Sestito, J.Mod.Opt., {\bf 51} (2004) 1063.
  \item[{\rm[18]}] G. Cassinelli, N. Zangh\'\i, Il nuovo Cimento B, {\bf 73} (1983) 237.
  \item[{\rm[19]}] M. Arndt, O. Nairtz, J. Voss-Andreae, C. Keller, G. Van Der Zouw, A. Zeilinger,
  Nature (London), {\bf 401} (1999) 680.
  \item[{\rm[20]}] T.J. Herzog, P.G. Kwiat, H. Weinfurter, A. Zeilinger, Phys.Rev.Lett.,
  {\bf 75} (1995) 3034.
  \item[{\rm[21]}] S. D\"urr, G. Rempe, Nature (London), {\bf 395} (1998) 33.
\end{description}

\end{document}